\documentclass[conference]{IEEEtran}
\usepackage{amsmath}
\usepackage{amsmath, amsthm, amssymb}
\usepackage[ansinew]{inputenc}
\usepackage{graphicx}
\usepackage{epstopdf}
\usepackage{cite}

\ifCLASSINFOpdf
\else
\fi
\begin{document}
\title{SINR-Threshold Scheduling with Binary Power Control for D2D Networks}
\author{\IEEEauthorblockN{Mehrdad Kiamari$^*$, Chenwei Wang$^{\dagger}$, A. Salman Avestimehr$^*$, and 	Haralabos Papadopoulos$^{\dagger}$}\\
\IEEEauthorblockA{$^*$ Department of Electrical Engineering, University of Southern California, Los Angeles, CA\\
$^{\dagger}$DOCOMO Innovations, Inc., Palo Alto, CA}}
\maketitle

\allowdisplaybreaks

\begin{abstract}
In this paper, we consider a device-to-device communication network in which $K$ transmitter-receiver pairs are sharing spectrum with each other. We propose a novel but simple binary scheduling scheme for this network to maximize the average sum rate of the pairs. According to the scheme, each receiver predicts its Signal-to-Interference-plus-Noise Ratio (SINR), assuming \emph{all} other user pairs are active, and compares it to a preassigned threshold to decide whether its corresponding transmitter to be activated or not. For our proposed scheme, the optimal threshold that maximizes the expected sum rate is obtained analytically for the two user-pair case and empirically in the general $K$ user-pair case. Simulation results reveal that our proposed SINR-threshold scheduling scheme outperforms ITLinQ \cite{navid}, FlashLinQ \cite{flash} and the method presented in \cite{G} in terms of the expected sum rate (network throughput). In addition, the computational complexity of the proposed scheme is $O(K)$, outperforming both ITLinQ and FlashLinQ that have $O(K^2)$ complexity requirements. Moreover, we also discuss the application of our proposed new scheme into an operator-assisted cellular D2D heterogeneous network.
\end{abstract}

%
\IEEEpeerreviewmaketitle

\section{Introduction}

Device-to-Device (D2D) refers to a wireless communication technology that enables direct communication between nearby devices. Establishing D2D communication links offers a means for offloading traffic over a cellular network. Once D2D communication links are connected and established, it is unnecessary to use the base station (BS) or other intermediate devices for data transmission. Driven by the prevalence of smart devices, the rapid growth of data load and internet of things, the forthcoming 5G aims at achieving significantly higher rates per unit area and significantly lower latency than the current LTE systems. Thus, the D2D technology is widely recognized, e.g., by 3GPP, as one of the key technologies for 5G.
To achieve the promised D2D benefits in 5G, several challenges need to be addressed, such as D2D user detection and identification \cite{D2D_detection}, D2D synchronization \cite{D2D_sync}, wireless resource management \cite{D2D_wrm}, power control and interference management \cite{Heath}, and efficient handovers between cellular access and D2D \cite{Shen}. In this paper, we focus on sharing the D2D channel among multiple user pairs and develop simple methods for power control and interference management.

\subsection{The Problem}
A D2D wireless network consisting of $K$ transmitter-receiver pairs is traditionally abstracted as a $K$-user interference channel. Although the capacity characterization of the $K$-user interference channel is still one of the most fundamental open problems in information theory, several recent discoveries have significantly advanced our understanding of the capacity limits, such as linear deterministic models \cite{Salman_Diggavi_Tse} and interference alignment \cite{Jafar_FTn}, almost all requiring global channel state information (CSI) be available at each user node.

As far as practical system designs are concerned, it is imperative to keep the encoding and decoding schemes as simple as possible. One such simple scheme in multiuser interference management is treating interference as noise (TIN) at the receiver. Recently, Geng \emph{et al.} in \cite{Geng} identified a TIN-optimality condition under which TIN is generalized degrees-of-freedom (GDoF) optimal and achieves the capacity region within a constant gap. However, the gap in the finite SNR regime could be large, particularly if the size of the network scales up. In addition, scheduling a small subset of users satisfying the TIN-optimality condition may not be necessarily the right metric for maximizing the expected sum rate in the network. Thus, the sum rate characterization of the $K$-user interference channel remains open. In particular, selecting a subset of user pairs which results in the highest sum rate has not been fully solved yet.
On this avenue, the problem we investigate in this work can be concisely expressed as follows: \emph{How to design a simple scheduling scheme with binary power control based on local (not global) CSI to achieve a higher sum rate than the state-of-the-art scheduling schemes?}

\subsection{The Prior Works}

On multiuser scheduling, there are a few recent works considering SINR-based methods, such as \cite{SINR-threshold} where a greedy algorithm was developed to improve the packet reception rate, but only an approximation of the optimal scheduling scheme is obtained in, e.g., \cite{SINR-threshold, SINR-threshold-minor, SINR-threshold-minor2}. Moreover, only a bound on SINR is derived and the ergodic sum rate is not characterized in \cite{SINR-threshold}. Besides SINR-based methods, a SNR-based scheme was introduced in \cite{Heath} and \cite{G} where whether to activate a link or not depends on the direct link information only. Since the interference links are not taken into account, activating a user-pair which causes strong interference to the others would reduce the sum rate of the network.

Of the vast amount of literature on the sum rate characterization of a D2D network, the most closely related to this work are references \cite{navid, flash, G}. Recently, Wu \emph{et al.} proposed a scheme referred to as FlashLinQ in \cite{flash}. It is a multiuser scheduling scheme where a user pair is scheduled if that user pair meets the FlashLinQ condition. The key of FlashLinQ is to activate a user-pair if the received SINR is strong enough and deactivate them if the transmitter generates too much interference.
Instead of testing the SINRs, Naderializadeh and Avestimehr proposed another scheme referred to as ITLinQ in \cite{navid}. Inspired by the TIN-optimality condition \cite{Geng}, they defined a new concept of information-theoretic independent sets (ITISs), which indicates the sets of links for which simultaneous communication and treating the interference from each other as noise is information-theoretically optimal (to within a constant gap). Based on this concept, they developed a new spectrum sharing mechanism, called information-theoretic link scheduling (ITLinQ), which at each time schedules those links that form an ITIS. It was demonstrated in \cite{navid} that ITLinQ can substantially improve over FlashLinQ. One great attribute of FlashLinQ and ITLinQ is that each user-pair is able to make the scheduling decision based on their local CSI. However, both schemes generate the active set incrementally, by checking the TIN type-optimality condition one user-pair at a time. Also, they require the presence of a central controller, which establishes a user-pair ordering that is to be used to generate the active user-pair set. As we elaborate in Section \ref{sec:scheme}, they both have computational complexity $O(K^2)$ and substantial overheads, rendering them inefficient at this stage for practical D2D implementation. Built upon ITLinQ, Yi {\em et al.} proposed the ITLinQ+ scheme in \cite{Yi_Caire} to achieve higher spectral efficiency via sophisticated non-binary power control.

To address the $O(K^2)$ computational complexity burden, Kerret \emph{et al.} developed a binary power control scheme for the $K=2$ interference channel in the framework of game theory. Specifically, each user-pair is selfishly turned on if the desired signal power is above a predetermined threshold. While the computational complexity of the method in \cite{G}, referred to as the SNR-based scheme, is reduced to $O(K)$, the sum rate performance is also worse than FlashLinQ and ITLinQ, due to the fact that the interference is not taken into account.

\subsection{Contributions}

In this paper, we propose a novel and quite simple SINR-threshold scheduling scheme with one-bit feedback and binary power control. Specifically, in a D2D network with $K$ transmitter-receiver pairs, the central controller of the network, e.g., a macro BS, broadcasts an \emph{optimal} threshold to each receiver. Each receiver tests if its SINR, assuming \emph{all} links are active, is above the given threshold based on its local CSI at the receiver (CSIR), different from the requirements of CSI at both the transmitter and the receiver for ITLinQ and FlashLinQ. Then each receiver feeds back this binary decision to its corresponding transmitter, so that the transmitter is either active in full power transmission or inactive. The predetermined threshold is a function of $K$ and it is broadcast to the users. In this work we derive the optimal threshold value (in terms of maximizing the expected sum rate) in closed form for $K=2$ user-pair case, assuming all channels are i.i.d. and Rayleigh distributed. We also empirically optimize the threshold  for general $K$-user case, involving more realistic wireless channel models, including pathloss and shadowing, etc.  As our simulation-based sum rate analysis reveals, the proposed scheme outperforms the ones in \cite{navid, flash, G}. For example, given the network with the parameters defined in \cite{navid} and $K = 800$, the simulation reveals that our scheme gains roughly 13$\%$, 75$\%$ and 128$\%$ sum rate increases compared to ITlinQ, FlashLinQ and the SNR-based scheme, respectively.


\subsection{Significance}

We next make a few important observations regarding the significance of this work. First, as mentioned earlier in this section, the sum rate characterization of D2D interference networks has attracted a lot of attention both in theory aiming at understanding the fundamental capacity limit via developing many ingenious ideas, and in practice with the goal of developing schemes as simple as possible to achieve the performance as higher as possible. Our work falls into the later category.

Second, our scheme outperforms in delivered sum rate of ITLinQ, FlashLinQ and the SNR-based scheme introduced in \cite{navid, flash, G}. We also set forth  two different scheduling approaches for improving user fairness at the cost of sum rate performance.  At the same time, fairness should be considered under the umbrella of cellular, as the availability of cellular access  provides an alternative to D2D for  serving certain user pairs. In light of this observation,  the sum rate appears to be a reasonable optimization metric for D2D networks.

Third, compared to FlashLinQ and ITLinQ, the computational complexity of our scheme is also significantly reduced from $O(K^2)$ to $O(K)$. And last but not least, from the network-wide operational perspective, our scheme only needs the central controller to broadcast a SINR threshold to all the users in the network according to the network size, which is much simpler than FlashLinQ and ITLinQ.

\section{System Model}

Consider a wireless D2D network where there are $K$ transmitter-receiver pairs $({\text{Tx}}_i,{\text{Rx}}_i)$ for $i=1,2,\cdots,K$, each node equipped with one antenna only, and each Tx$_i$ aims to send one independent message to its intended Rx$_i$. Formulated as a $K$-user interference channel, the system input-output relationship can be characterized by the following equation:
\small
\begin{eqnarray}
y_i(t) = \sum_{j=1}^K h_{ij}x_j(t) + z_i(t),~~~~i\in \mathcal{K}\triangleq \{1,2,\cdots,K\},\label{eqn:channelmodel}
\end{eqnarray}
\normalsize
where $y_i(t)$ represents the received signal at Rx$_i$ at time $t$; $x_j(t)$ represents the transmitted signal from Tx$_j$ at time $t$ satisfying the power constraint $\mathbb{E}[\frac{1}{T}\sum_{t=1}^T|x_j(t)|^2]\leq P$; $z_i(t)\sim \mathcal{CN}(0,N_o)$ denotes the i.i.d. additive white Gaussian noise (AWGN); $h_{ji}$ represents the complex channel coefficient from Tx$_i$ to Rx$_j$. We assume that the channel coefficients $h_{ij}$'s are drawn from a continuous distribution, which will be specified in detail in Section \ref{sec:K2} and Section \ref{sec:simulation}, and they stay constant during the entire transmission once they are drawn. We also assume that all the transmitter-receiver pairs share a common wireless medium, i.e., given the same chunk of frequency-time resources, we need to deal with considerable multiuser interference. In this work, we assume that local CSI is available at each receiver, i.e., $\{h_{ij}\}_{j=1}^K$ is known to Rx$_i$. In practice, this can be accomplished by training the channels via sending a pilot symbol from each transmitter to all receivers at a time and allowing each receiver for channel estimation.

Next, we introduce the formal definition of binary power control aiming at maximizing the expected sum rate based on the local CSI at each receiver.

\subsection{Binary Power Control}

Consider the Best Response (BR) approach introduced in \cite{G}, and define the \emph{binary power control} function at Tx$_i$ as:
\begin{equation}
p_i(f_i):~f_i\longrightarrow ~~\{0,P\},~~~~i\in\mathcal{K},
\end{equation}
where $f_i\in \mathbb{R}^+$ represents the feedback function from Rx$_i$ to Tx$_i$, and the optimal power control scheme can be written as
\small
\begin{eqnarray}\label{eq111opti}
(p_1^*,\cdots,p_K^*)=\text{arg} \max_{(p_1,\cdots,p_K)\subset \mathcal{P}} \mathbb{E}[R_{\Sigma}(p_1(f_1),\cdots,p_K(f_K))]\!\!
\end{eqnarray}
\normalsize
where $\mathcal{P}$ indicates the set corresponding to all possible power control functions $(p_1,\cdots,p_K)$.

This optimization problem is a typical \emph{Team Decision} problem \cite{G12} due to the fact that each transmitter aims to maximize the ergodic sum rate based on their individual channel information. To simplify this optimization problem, a modified version of BR for the $K=2$ case was defined in \cite{G} and is rewritten as follows:

{\it Definition 1 \cite{G}:} $(p_1^{\textrm{BR}},p_2^{\textrm{BR}})$ are the best response power control functions if they satisfy
\begin{equation}
\begin{aligned}
p_1^{\textrm{BR}} &\in \text{arg} \max_{(p_1,p_2^{\textrm{BR}})\subset \mathcal{P}} \mathbb{E}[R_{\Sigma}(p_1(f_1),p_2^{\textrm{BR}}(f_2))],\\
p_2^{\textrm{BR}} &\in \text{arg} \max_{(p_1^{\textrm{BR}},p_2)\subset \mathcal{P}} \mathbb{E}[R_{\Sigma}(p_1^{\textrm{BR}}(f_1),p_2(f_2))].
\end{aligned}
\end{equation}

It should be noted that the best response defined in \cite{G} is only based on the CSI of the direct link, i.e., without considering interference, and the optimal power control functions $(p_1,p_2)$ are also the best response power control functions \cite{G}.

\section{The Proposed SINR-Threshold Scheduling Scheme}\label{sec:scheme}

Since resolving the optimization problem (\ref{eq111opti}) is complicated, we focus on a smaller functional space called a \emph{threshold function}. In addition, to reduce the feedback cost, we impose the constraint that each feedback function $f_i$ has up to one bit only, i.e., its entropy is upper bounded by one. In this section, we propose a novel and very simple SINR-threshold scheduling scheme based on local CSIR for the binary power control problem. The scheme consists of the learning phase, the scheduling phase and the communication phase.

\subsection{The Learning Phase}

In this phase, each Rx$_i$ acquires its local CSI of $\{h_{ij}\}_{j=1}^K$, by reading the pilot symbol sent by one transmitter every time. Then, all users predict their SINRs by assuming all links to be active. The predicted SINR at Rx$_i$ is calculated as follows:
\begin{eqnarray}\label{eqn:opt_problem}
\textrm{SINR}_i \triangleq \frac{|h_{ii}|^2}{\sum_{j\neq i}|h_{ij}|^2+N_o/P},~~~~i\in\mathcal{K}.
\end{eqnarray}

\subsection{The Scheduling Phase}

After completing the learning phase, each Rx$_i$ compares its SINR with the threshold $\gamma$ which is prior assigned by, e.g., the macro-cell BS to each receiver. Then each receiver {\em distributively} determines whether to activate its corresponding transmitter or not according to the following definition.

{\it Definition 2 (The SINR-Threshold Scheduling with Binary Power Control):} We choose the feedback function as an indicator function $f_i\triangleq{\bf 1}(\textrm{SINR}_i-\gamma)$, and a functional subset $\mathcal{G} \subset \mathcal{P}$ is defined as:
\small
\begin{equation}
\mathcal{G}\!\triangleq\!\{(p_1,\!\cdots\!,p_K)|(p_1,\!\cdots\!,p_K) \in \mathcal{P},~ p_i(f_i)=P{\bf 1}(f_i),~i\in\mathcal{K}\}
\end{equation}
\normalsize
where ${\bf 1}(x)$ is the indicator function with ${\bf 1}(x)=1$ if $x\geq 0$ and ${\bf 1}(x)=0$ if $x< 0$.

After the value of $f_i$ is fed from Rx$_i$ back to Tx$_i$\footnote{The availability of feedback can be accomplished via either the feedback channel from the receiver to its transmitter or the help of the macro-cell BS through uplink and then downlink cellular transmission.}, the transmitter Tx$_i$ determines its status in the next phase of data transmission via the mapping function $p_i(f_i)$. Specifically, Tx$_i$ will be activated with full power if the received SINR at its corresponding receiver is above its preassigned threshold; Otherwise Tx$_i$ will keep silent. Thus, only the user pairs in the subset ${\mathcal S}\triangleq\{i | f_i=1\}$ are scheduled for transmission given the channel realizations $H\triangleq \{h_{ij}|i,j\in\mathcal{K}\}$.

{\it Remark 1:} Since the feedback $f_i$ is an indicator function, it contains up to one bit only. Also, the SINR threshold $\gamma$ depends on both the desired signal power and interference strength. In contrast, the threshold in \cite{G} depends on the CSI of the desired link only (i.e., the SNR). Thus, one can expect that the sum rate of our scheme outperforms that in \cite{G}.

\subsection{The Communication Phase}

After determining the subset $\mathcal S$ of user pairs for scheduling, the scheduled user-pair $i$ can achieve the following rate\footnote{Even without CSIT, the rate in (\ref{eqn:rate}) can be achieved arbitrarily closely with HARQ at the cost of decoding delay, if the traffic is not delay sensitive.} by using TIN as the decoding scheme:
\small
\begin{eqnarray}\label{eqn:rate}
R_i(H,\gamma)&\!\!\!\!=\!\!\!\!&\log\Big(1+\frac{|h_{ii}|^2}{\sum_{j\in\mathcal{S},j\neq i}|h_{ij}|^2+N_o/P}\Big),~~\forall i \in \mathcal S.\ \ \
\end{eqnarray}
\normalsize
Thus, the following sum rate $R_{\Sigma}(H,\gamma)$ and the ergodic sum rate $\bar{R}_{\Sigma}(\gamma)$ are achievable:
\small
\begin{eqnarray}
R_{\Sigma}(H,\gamma)&\!\!=\!\!&\sum_{i\in {\mathcal S}}{R_i(H,\gamma)},\\
\bar{R}_{\Sigma}(\gamma)&\!\!=\!\!&\mathbb{E}_H[R_{\Sigma}(H,\gamma)].
\end{eqnarray}
\normalsize
Finally, the optimal sum rate of the network offered by our scheme is characterized by the optimization problem:
\begin{eqnarray}\label{eqn:opt_problem}
\max~~\bar{R}_{\Sigma}(\gamma)~~~~~~\textrm{s.t.}~~\gamma\in[0,+\infty).
\end{eqnarray}

{\it Remark 2:} Consider the computational complexity of our scheme, ITLinQ and FlashLinQ. Since they all require local CSIR which can be obtained with the linear complexity, we do not consider the cost of acquiring CSI and the corresponding SINR values, but consider the scheduling decision making only. In our scheme, each receiver only needs to compare its SINR to the given threshold for once, so that the total computational complexity of $K$ user-pairs is given by $O(K)$. In contrast, in both ITLinQ and FlashLinQ, when we check whether or not to add a new user-pair into the subset including user pairs chosen for scheduling, we have to check their corresponding conditions at {\em each} transmitter and {\em each} receiver in that subset, and schedule the new transmitter-receiver pair if they both satisfy the condition. This results in the total complexity up to $6+9+\cdots+3K=1.5(K+2)(K-1)$, i.e., $O(K^2)$. Thus, the computational complexity of our proposed scheme outperforms both ITLinQ and FlashLinQ.

Based on the proposed scheme, it can be seen that to achieve the maximum sum rate in (\ref{eqn:opt_problem}), we need to find out the value of the corresponding optimal threshold $\gamma$. To do this, we will begin with deriving the closed form expression of the expected sum rate for the $K=2$ case in Section \ref{sec:K2}, and then proceed to evaluating the expected sum rate for the general $K$-user case via simulation in Section \ref{sec:simulation}, followed by a few discussions.

\section{Optimal Threshold Selection}\label{sec:K2}

In this section, we first analytically derive the optimal threshold for $K=2$. For the general $K$-user case, due to the difficulty of expressing the closed form ergodic rate in terms of the threshold, the optimal threshold can be numerically obtained based on search over the effective SINR regime. 


For $K=2$, to simplify the problem (\ref{eqn:opt_problem}) as much as possible, we assume the channel coefficients $h_{ji}$'s are i.i.d. and follow the Rayleigh distribution $h_{ji}\sim \mathcal{CN}(0,1)$, and the noise is normalized to unit power $N_o=1$, the same assumption as in \cite{G}. Under this setting, we have the following lemma:

{\it Lemma 1:} Suppose $p_i^{\gamma_i}, i=1,2$ are power control functions, then there exists a pair $(\gamma_1^*,\gamma_2^*) $ so that $(p_1^*,p_2^*)=(p_1^{{\gamma_1}^*},p_2^{{\gamma_2}^*})$.

\begin{proof} The proof follows directly from \cite{G} where we replace the SNR threshold with the SINR threshold in this paper. \end{proof}

To maximize the ergodic sum rate in (\ref{eqn:opt_problem}), we will develop the analytical expression of the sum rate $\bar{R}_{\Sigma}$ in (\ref{eqn:opt_problem}) for $K=2$, and then find out the optimal threshold. For brevity, we rewrite the mapping function $p_i=X_i$ as follows:
\begin{equation}\label{eq2}
X_i \triangleq
\left\{
	\begin{array}{ll}
		1  & \mbox{if } g_{ii} \geq \gamma(g_{i{\bar i}}+{1\over P}) \\
		0 & \mbox{if } \text{else}
	\end{array}
\right. ~~i=1,2~\textrm{and}~\bar i\triangleq 3-i
\end{equation}
where $g_{ij}\triangleq|h_{ij}|^2$ for $\forall i,j$ follow the exponential distribution with mean one. Hence, the ergodic sum rate is given by:
\small
\begin{eqnarray}
\bar{R}_{\Sigma}(\gamma)&\!\!\!\!=\!\!\!\!&\mathbb{E}[R_1+R_2]=2\mathbb{E}[R_1]\\
&\!\!\!\!=\!\!\!\!&2\mathbb{E}\left[\log\left(1+{{g_{11}X_1P}\over{{g_{12}X_2P+1}}}\right)\right]\label{eq30}\\
&\!\!\!\!=\!\!\!\!&2\mathbb{E}_{g_{12},X_2}\!\!\!\left[\int_{\gamma(g_{12}+{1\over P})}^{+\infty}{\!\!\!\!\log\!\left(1\!\!+\!\!{{g_{11}}\over{{g_{12}X_2\!+\!\!{1 \over P}}}}\right)\!e^{\!-g_{11}\!}dg_{11}}\right]\ \ \ \ \ \label{eq3}
\end{eqnarray}
\normalsize
where the first equal sign in (\ref{eq30}) is due to the fact that the rates for the two users are statistically the same. Since $X_2$ is a binary number, we can continue to rewrite $\bar{R}_{\Sigma}(\gamma)$ as follows

\small
\begin{eqnarray}
\bar{R}_{\Sigma}&\!\!\!\!=\!\!\!\!&2\mathbb{E}_{g_{12},X_2}\left[(1-X_2)\!\!\int_{\gamma(g_{12}+{1\over P})}^{+\infty}\!\!\!\!{\log(1+{{g_{11}}P})e^{-g_{11}}dg_{11}}\right]\notag\\
&\!\!\!\!\!\!\!\!&+2\mathbb{E}_{g_{12},X_2}\left[X_2\int_{\gamma(g_{12}+{1\over P})}^{+\infty}\!\!\!\!{\log\left(\!\!1\!+\!{{g_{11}}\over{{g_{12}+{1 \over P}}}}\!\!\right)\!\!e^{-g_{11}}dg_{11}}\right]\\
&\!\!\!\!=\!\!\!\!&2\mathbb{E}_{X_2}\!\!\left[(1-X_2)\!\!\int_{0}^{+\infty}\!\!\!\!{\log(1\!+\!\gamma\!+\!\gamma P g_{12})e^{\!-\!(\gamma+1)g_{12}\!-\!{{\gamma}\over P}}dg_{12}}\right]\ \ \ \ \notag\\
&\!\!\!\!\!+\!\!\!\!\!&2\mathbb{E}_{X_2}\left[(1-X_2)\int_{0}^{+\infty}\!\!\!\!{E_i\left(\gamma g_{12}+{{\gamma+1}\over{P}}\right)e^{{1 \over P}-g_{12}}dg_{12}}\right]\notag\\
&\!\!\!\!\!\!\!\!&+2\mathbb{E}_{X_2}\left[X_2\int_{0}^{+\infty}\!\!\!\!{\log(1+\gamma)e^{-(\gamma+1)g_{12}-{{\gamma}\over P}}dg_{12}}\right]\notag\\
&\!\!\!\!\!\!\!\!&+2\mathbb{E}_{X_2}\left[X_2\int_{0}^{+\infty}\!\!\!\!{E_i\left((1+\gamma) (g_{12}+{1\over{P}})\right)e^{{1 \over P}}dg_{12}}\right]\label{eq4}
\end{eqnarray}
\normalsize
where $E_i(x)=\int_{x}^{\infty}{{{e^{-t}}\over{t}}dt}$. Furthermore, we define:
\small
\begin{equation}
\begin{aligned}
A_1(\gamma)&\triangleq {{e^{{-\gamma}\over{P}}}\over{1+\gamma}}\left(\log(1+\gamma)+e^{{{{(1+\gamma)}^2}\over{\gamma P}}}E_i\Big({{{(1+\gamma)}^2}\over{\gamma P}}\Big)\right),
\\A_2(\gamma) &\triangleq e^{{1\over{P}}}E_i\Big({{{1+\gamma}}\over{P}}\Big)-e^{{{2\gamma+1}\over{\gamma P}}}E_i\Big({{{(1+\gamma)}^2}\over{\gamma P}}\Big),
\\A_3(\gamma) &\triangleq{{e^{{-\gamma}\over{P}}}\over{1+\gamma}}\log(1+\gamma),
\\A_4(\gamma) &\triangleq {e^{{-\gamma}\over{P}}}  \left( {1\over{1+\gamma}}-{1\over P}e^{{{{(1+\gamma)}}\over{P}}}E_i\Big({{{1+\gamma}}\over{P}} \Big)\right).
\end{aligned}
\end{equation}
\normalsize
By substituting $\mathbb{E}_{X_2}[X_2]=\textrm{Prob}(X_2=1)={{e^{-{{\gamma}\over{P}}}}\over{1+\gamma}}$ into (\ref{eq4}), $R_{\Sigma}(\gamma)$ can be further written in the following  compact form:
\small
\begin{eqnarray}
\bar{R}_{\Sigma}(\gamma)=2\Big(1-{e^{-{\gamma}\over{P}}\over{1+\gamma}}\Big)(A_1(\gamma)+A_2(\gamma))~~~~~~~~~~~~\notag\\
~~~~~~~~~~~~~~~ +{2e^{-{\gamma}\over{P}}\over{1+\gamma}}(A_3(\gamma)+A_4(\gamma)).
\end{eqnarray}
\normalsize
Note that given the power $P$, $\bar{R}_{\Sigma}(\gamma)$ depends on the threshold $\gamma$ only. To find the optimal threshold $\gamma^*$, we take the first derivative of $\bar{R}_{\Sigma}(\gamma)$ w.r.t. $\gamma$ and solve $d{\bar{R}_{\Sigma}(\gamma)}/d\gamma|_{\gamma=\gamma^*}=0$. While the closed form expression of $\gamma^*$ is difficult to derive, it can be solved by applying many numerical algorithms such as the Newton method and the gradient descending algorithm, and the result can be easily verified via simulation.


For the general $K$-user case, we resort to simulations. In particular, we use $\frac{1}{T_s}\sum_{t=1}^{T_s} \bar{R}_{\Sigma}(H_t,\gamma)$ to approximate $\bar{R}_{\Sigma}(\gamma)$ by letting $T_s$ be large enough where $T_s$ is the total number of randomly drawn channel realization sets and $H_t$ represents the $t^{th}$ set for $t=1,2,\cdots,T_s$.
In addition, when the user distance, channel fading and shadowing are taken into account, the objective function $\bar{R}_{\Sigma}(\gamma)$ is not convex and even not continuous w.r.t. $\gamma$ due to use of the indicator function. Thus, we need to develop a new method to find the {\em global} optimal $\gamma^*$ in Problem (\ref{eqn:opt_problem}). We state our result as follows.

{\it Lemma 2:} There exists an optimal $\gamma^*$ so that $\bar{R}_{\Sigma}(\gamma)\leq \bar{R}_{\Sigma}(\gamma^*)$, and $\gamma^*$ can be obtained in $O(KT_s)$.

\begin{proof} We show first the converse and then the achievability.

({\it Converse}) Recall the scheduling policy in Definition 2 where user-pairs are scheduled only if their SINRs are no less than $\gamma$. That is, once a realization set $H$ is given, the predicted SINR at each receiver is determined. Let us sort the SINRs as SINR$_{\pi(1)}<$ SINR$_{\pi(2)}<\cdots<$ SINR$_{\pi(K)}$ where $\pi(k)$ maps to the index of the user pair which has the $k^{th}$ smallest value of SINR. It can be easily seen that for $\gamma\in[\textrm{SINR}_{\pi(k-1)},\textrm{SINR}_{\pi(k)})$, $R_{\Sigma}(H,\gamma)$ is a continuous function of $\gamma$. Thus, given $T_s$ channel realization sets $\{H_t\}_{t=1}^{T_s}$, $\bar{R}_{\Sigma}(\gamma)$ is still piecewise continuous of $\gamma$. Also, note that the effective SINR regime to choose $\gamma$ from is bounded by the smallest and the largest values of SINRs obtained from $K$ receivers over $T_s$ times. For a piecewise continuous function on a bounded regime, $\bar{R}_{\Sigma}(\gamma)$ is also upper bounded.

({\it Achievability}) For $\gamma\in[\textrm{SINR}_{\pi(k-1)},\textrm{SINR}_{\pi(k)})$, $R_{\Sigma}(H,\gamma)$ is not only continuous but also piecewise constant valued w.r.t. $\gamma$. Hence, when $\gamma$ increases, $R_{\Sigma}(H,\gamma)$ can achieve a new value only when $\gamma$ hits an SINR$_i$ value for $i\in \mathcal K$. Thus, given $T_s$ channel realization sets $\{H_t\}_{t=1}^{T_s}$, we only need to calculate $\bar{R}_{\Sigma}(\gamma)$ when $\gamma=\textrm{SINR}_{\pi(k)}^{H_t}$, i.e., the SINRs when $H_t$ is given. Since there are $K$ users and $T_s$ channel realization sets for simulation, the optimal threshold $\gamma^*$ can be found out from the corresponding largest $\bar{R}_{\Sigma}(\textrm{SINR}_{\pi(k)}^{H_t})$ in $O(KT_s)$.
\end{proof}

\section{Performance Analysis}\label{sec:simulation}

\begin{figure}[!b]\vspace{-0.2in}
\centering
\includegraphics[width=0.2\textwidth]{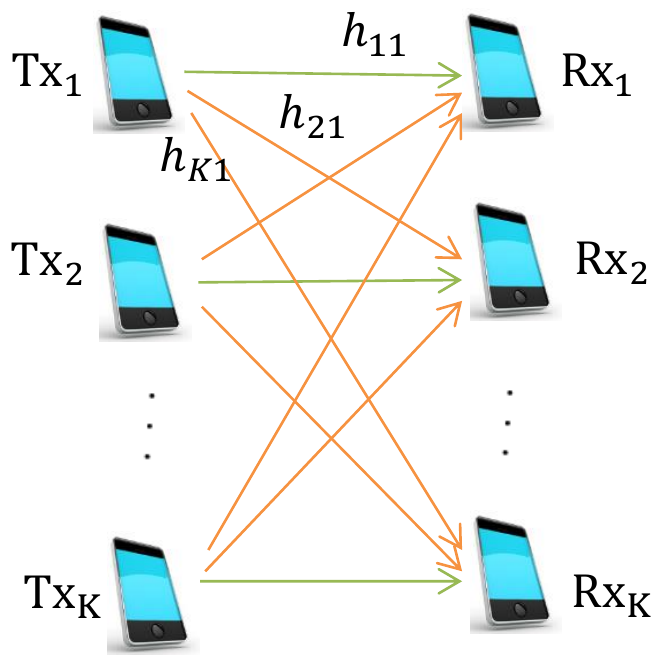}\vspace{-0.15in}
\caption{The system model of $K$-user device-to-device network where ${Tx}_i$ and ${Rx}_i$ represent $i$th transmitter and $i$th receiver respectively.}
\label{system_model}
\end{figure}

In this section, we will numerically compare the performance of the SINR-threshold scheduling scheme proposed in this paper with ITLinQ, Fair ITLinQ, FlashLinQ and the SNR-based scheduling scheme studied in \cite{navid, flash, G} in terms of the sum ergodic rate and the CDFs of the user rates for the general $K$-user D2D network depicted in Fig. \ref{system_model}. Note that choosing these two metrics is consistent with \cite{navid}-\cite{flash} for a fair comparison. As far as the parameter setting is concerned, we consider the size of the area is given by $1000$m$\times 1000$m, 
and there are $K$ transmitter-receiver pairs randomly dropped. For fair performance comparison, the same as in \cite{navid}, we consider the distance between each transmitter and its receiver is drawn from the uniform distribution $U[2\textrm{m},65\textrm{m}]$. Furthermore, we assume that the maximum transmit power for each user is given by $P=20$ dBm, and the noise power spectral density is $N_o=-184$ $\text{dBm}/\text{Hz}$ respectively. The noise figure is set to $7$ dB, and the antenna gain per device is assumed to be $-2.5$ dB. Moreover, we assume that the carrier frequency is $2.4$ GHz and the bandwidth is $5$ MHz. In addition, we assume that all antennas heights are set to be $1.5$m. As far as path loss is concerned, the transmission power loss in dB value at distance $d$ can be modeled the same as \cite{navid} as follows
\begin{equation}
L = L_{bp}+6+\left\{
	\begin{array}{ll}
		20 \log({d\over{R_{bp}}})  & \mbox{if } d\leq R_{bp} \\
		40 \log({d\over{R_{bp}}})  & \mbox{if } d> R_{bp}
	\end{array}
\right.
\end{equation}
where $R_{bp} \triangleq {{4h_bh_m}\over{\lambda}}$, $L_{bp} \triangleq \bigg|20\log\Big({{{\lambda}^2}\over{8\pi h_bh_m}}\Big)\bigg|$,
and $h_b$, $h_m$, $\lambda$ represent the BS antenna height, the mobile antenna height, and the transmission wavelength, respectively.

\subsection{The Expected Sum Rate Comparison}

\begin{figure}[!t]
\centering
\includegraphics[width=90 mm]{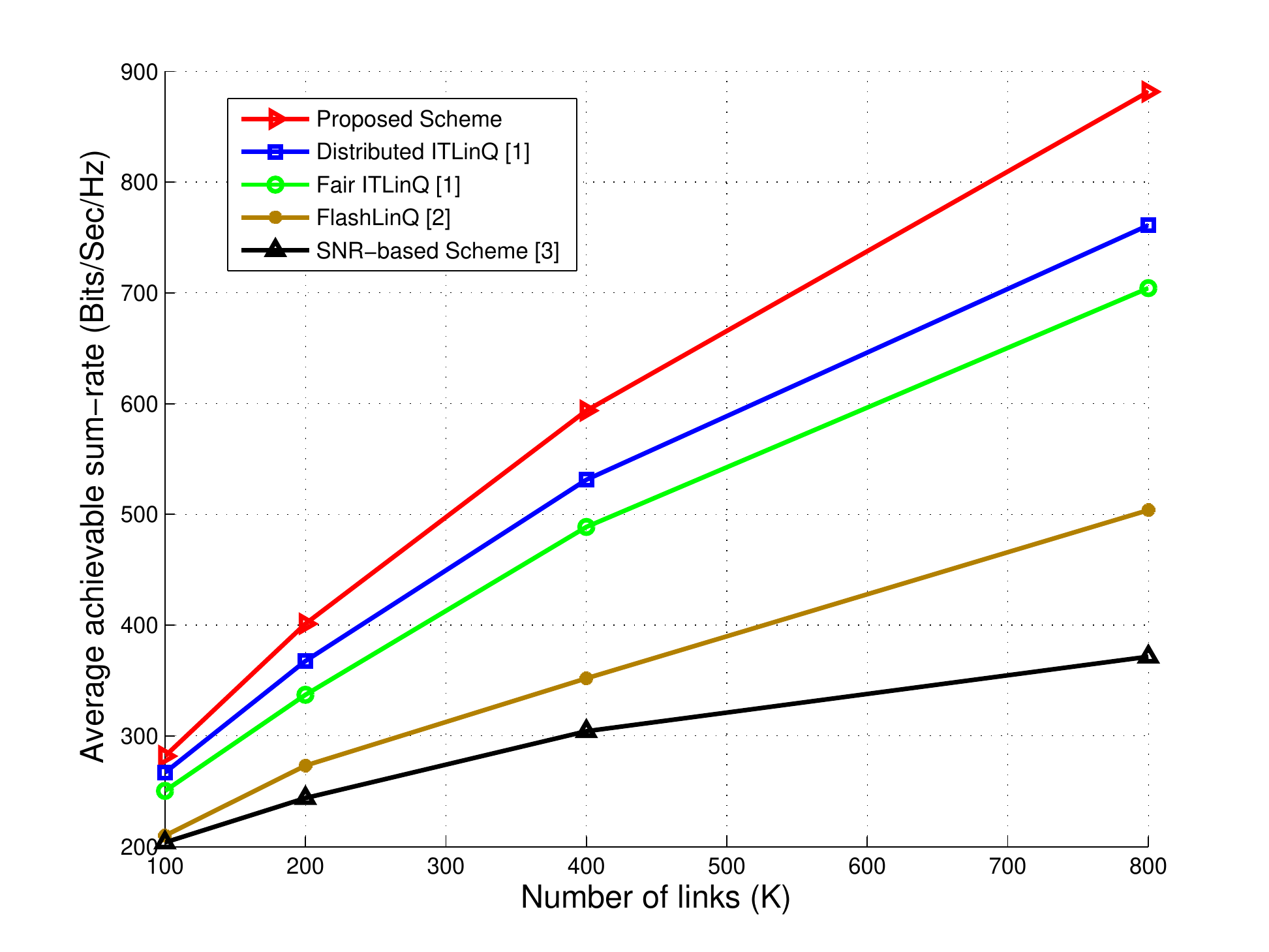}\vspace{-0.15in}
\caption{The average sum rate of the proposed scheme compared to ITLinQ, Fair ITLinQ, FlashLinQ, and the SNR-based scheme.}\label{su1000}
\end{figure}

Fig. \ref{su1000} demonstrates the average sum rate of our proposed scheme, ITLinQ \cite{navid}, FairITLinQ \cite{navid}, FlashLinQ \cite{flash} and the SNR-based scheduling scheme \cite{G}. Regarding the ITLinQ scheme, to have a fair comparison, we consider the optimal value for parameter $\eta$ of ITLinQ (which captures the interplay between the desired signal power and strongest interference power, see \cite{navid}). For FairITLinQ, we consider following setting of parameters: $\textrm{SNR}_{th}=110$ dB, $M=25$ dB, $\bar M=20$ dB and $\bar \eta=0.6$ (see \cite{navid}). As far as FlashLinQ is concerned, we follow similar implementation considered in \cite{flash} with $\gamma_{TX}=\gamma_{RX}=9$ dB. As Fig. \ref{su1000} shows, the sum rate of our proposed scheme always outperforms all the other schemes. In particular, our scheme can gain roughly $13\%$, $22\%$, $75\%$ and $128\%$ rate increases compared to ITlinQ \cite{navid}, FairITLinQ \cite{navid}, FlashLinQ \cite{flash} and SNR-based scheme \cite{G}, respectively, in the setting with $K=800$.

\begin{figure}[!t]
\centering
\includegraphics[width=90 mm]{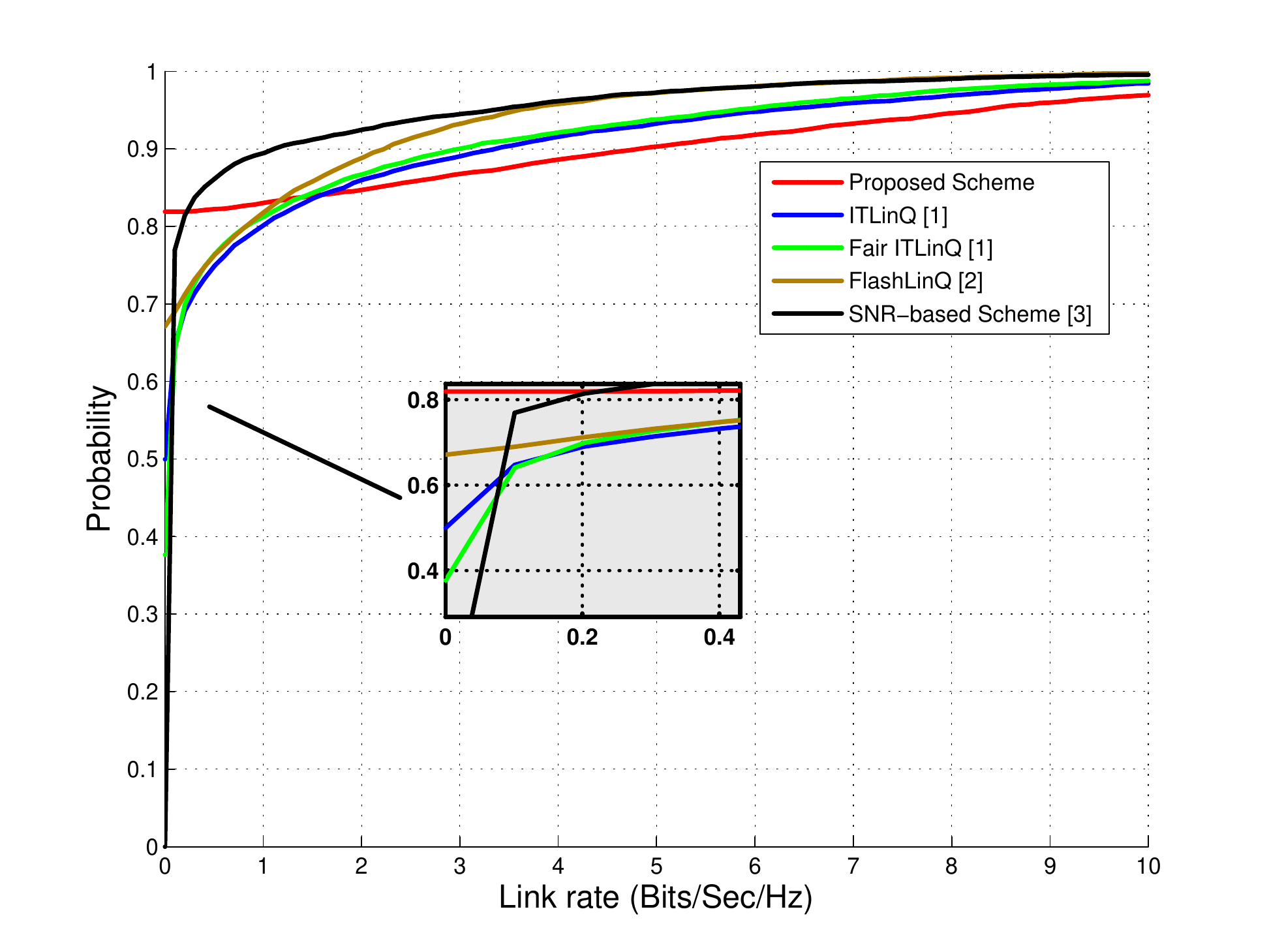}\vspace{-0.15in}
\caption{The CDF of user rates in the proposed scheme compared to ITLinQ, Fair ITLinQ, FlashLinQ, and the SNR-based scheme for $K=800$ users.}\label{c1000}
\end{figure}

Fig. \ref{c1000} provides the corresponding CDF curves of all the schemes shown in Fig. \ref{su1000}.
It can be seen the sum rate gain of our scheme is at the cost of user fairness. For example, the proposed scheme deactivates roughly $81\%$ of user pairs, significantly higher than $50\%$ of ITLinQ and $40\%$ of FlashLinQ. However, as mentioned earlier in this paper, fairness in a D2D network \emph{only} should not be a fatal issue, because fairness should be considered under the umbrella of cellular heterogeneous networks.

\subsection{Discussion: Trade Sum Rates for the User Fairness}

Although the fairness is not our focus in this work, we still provide two potential promising approaches which are able to improve the user fairness at the cost of, unsurprisingly, the sum rates. We present the two approaches as follows.

\subsubsection{Modified Proposed Scheme 1 (Enforcing A Portion of Links to be Activated)}

The idea is to enforce partial user links to be activated, no matter if their received SINRs are strong enough or not. For the remaining user pairs, we directly apply our scheme considering the interference generated by the enforced user pairs simultaneously.

Consider an example where $10\%$ of user links are enforced to be activated in a Round Robin manner. The average sum rate and the user rate CDF curves for $K=800$ are shown in Fig. \ref{fig:modified_sum} and Fig. \ref{fig:modified_cdf}, compared to our originally proposed scheme and ITLinQ. It can be easily seen that the number of inactive user links reduces from $81\%$ to $0$ because each user has a chance to talk due to Round Robin scheduling. Also, although we lose a certain sum rate compared to our proposed original scheme, we can still achieve higher sum rate than other schemes in \cite{navid, flash, G} while keeping the computational complexity benefit.



\subsubsection{Modified Proposed Scheme 2 (Reducing Threshold to Activate More Links)}

The idea is to reduce the threshold from the optimal value by a certain level, so that more user pairs could be scheduled.

\begin{figure}[!t]
\centering
\includegraphics[width=90 mm]{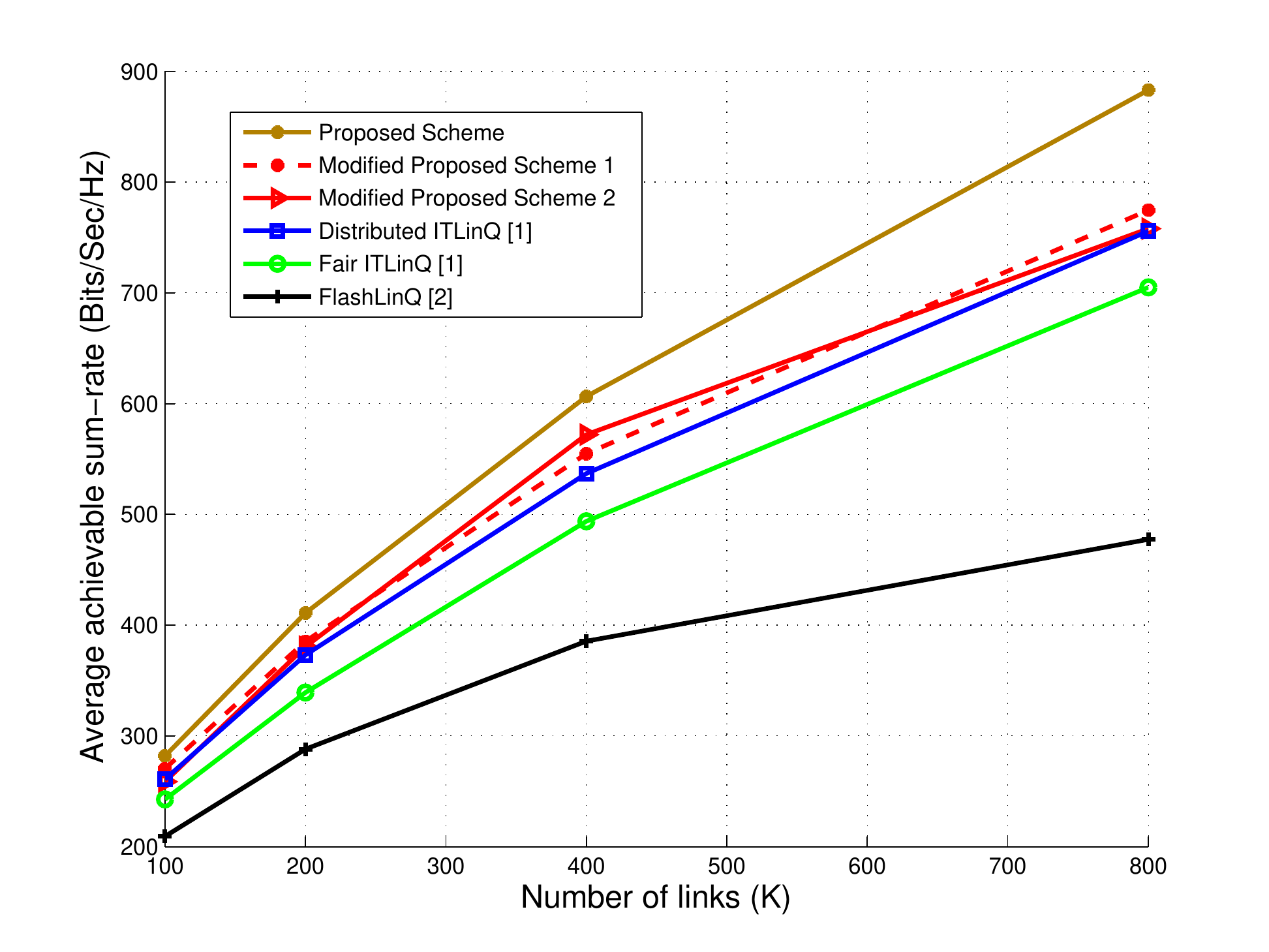}\vspace{-0.15in}
\caption{The average sum rate of the modified schemes 1 and 2 compared to the original proposed scheme, ITLinQ, Fair ITLinQ, FlashLinQ, and the SNR-based scheme.}
\label{fig:modified_sum}\vspace{-0.15in}
\end{figure}

\begin{figure}[!t]
\centering
\includegraphics[width=90 mm]{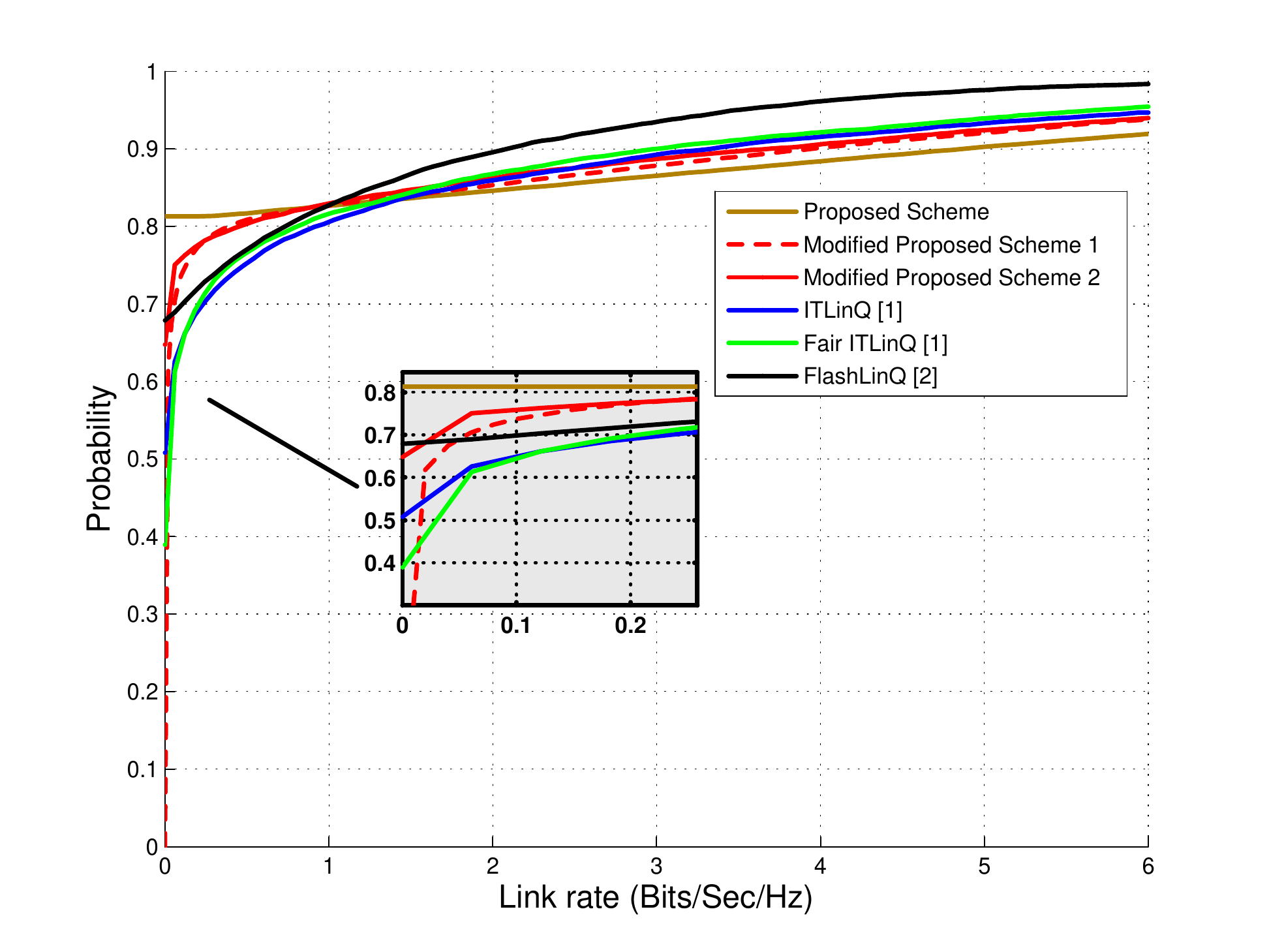}\vspace{-0.15in}
\caption{The CDF user rates of the modified scheme 1 and 2 compared to the original proposed scheme, ITLinQ, Fair ITLinQ, FlashLinQ, and the SNR-based scheme for $K=800$ users.}
\label{fig:modified_cdf}
\end{figure}

Consider $\gamma=\gamma^*-\gamma_{v}$ as an example where $\gamma^*$ is the optimal threshold according to our original scheme, $\gamma_{v}$ is a constant to be designed, and $\gamma$ is the threshold that we apply in this modified scheme. The average sum rate and the user rates CDF curves for $K=800$ are shown in Fig. \ref{fig:modified_sum} and Fig. \ref{fig:modified_cdf} and compared to ITLinQ and FlashLinQ. Since decreasing the threshold results in more fairness at the cost the sum rate, the offset $\gamma_{v}$ can be tuned to gain a tradeoff between the sum rate and the user fairness. Under our simulation setting and consider $K=800$ as an example.
When we choose $\gamma_v=0.45$, Fig. \ref{fig:modified_cdf} reveals that the user fairness of this modified scheme can be also improved compared to the original scheme, and we still have the sum rate gain, roughly $4.2\%$ more than ITlinQ and the computational complexity benefit.

{\it Remark 3:} Observations of Fig. \ref{fig:modified_cdf} reveal that when user rates are very close to zero, the modified propose scheme 1 outperforms both Fair ITLinQ and FlashLinQ in terms of fairness. On the other hand, from the {\em operational perspective}, {\em all} the schemes identified in Fig. \ref{fig:modified_cdf} are not very efficient in terms of fairness, since roughly $2/3$ of the users cannot achieve rate significantly larger than zero. One possible method is to schedule a subset of users instead of considering all users (corresponding activity fraction equal to one) during each slot.


\subsection{Thresholds for Network Operation}

\begin{figure}[!t]
\centering
\includegraphics[width=73 mm]{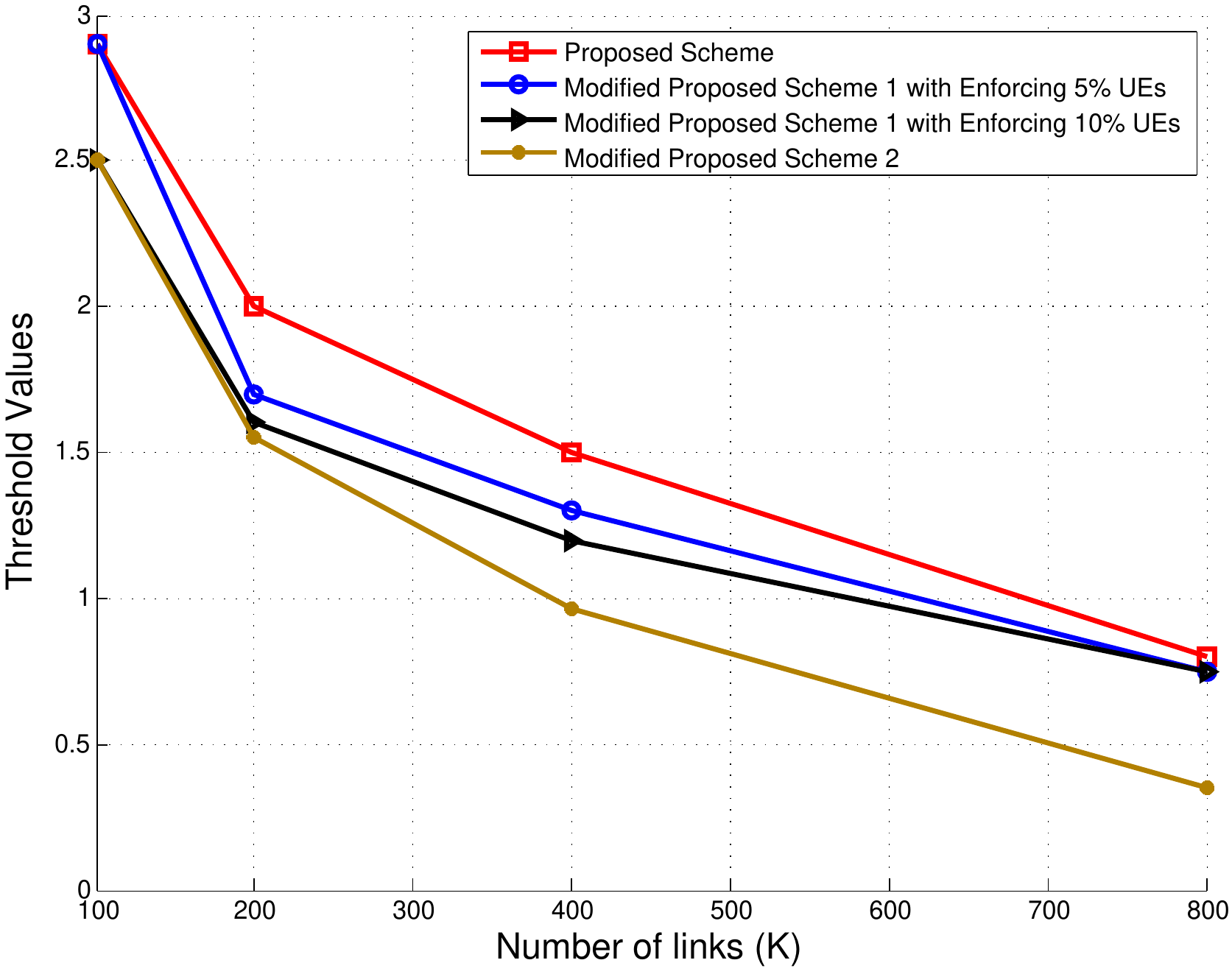}\vspace{-0.15in}
\caption{The predetermined threshold values assigned by the macro BS.}\label{fig:R1000_threshold}
\end{figure}

Since the optimal threshold $\gamma^*$ developed from Section \ref{sec:K2} is a function of the network size $K$, it can be predetermined in the network design. Based on our approaches in Section \ref{sec:K2}, Fig. \ref{fig:R1000_threshold} demonstrates the optimal threshold values for different $K$ values. When the D2D network is being operated, the macro BS only needs to first find the threshold values according to the network size, and then broadcast this value to all users in the D2D network. Meanwhile, the thresholds of the modified schemes 1 and 2 introduced above are also shown in Fig. \ref{fig:R1000_threshold}.

\subsection{Cellular-assisted D2D Network Operation}\label{sec:hetnet}

Observations of the user fairness result shown in Section \ref{sec:simulation} reveal that \emph{all} the schemes identified in Fig. \ref{fig:modified_cdf} are not efficient. However, as mentioned earlier in this paper, fairness by itself is not a fatal issue due to the fact that fairness in D2D networks should be considered under the umbrella of cellular heterogeneous networks, i.e., each user-pair can be activated in both either D2D or cellular bands. Due to the space limitation and since the user fairness is not the focus of this work, we will briefly discuss the cellular-assisted D2D heterogeneous networks, and defer the detailed analysis into the future work.

We consider a simple example of a cellular-assisted D2D heterogeneous network where in the D2D network, the rate of user $k$ is given by $\bar{R}_k(\gamma)=\mathbb{E}_H(R_k(H,\gamma))$. On the other hand, in the cellular network, there are $B$ cells, each with one BS in the cell center, and each BS schedules $K_c$ user at each time. After BS-user association and load balancing, each scheduled user is associated with one BS only. Meanwhile, we assume that the nominal user rate provided by the cellular network is given by $R_{c}$, implying that the ergodic user rate provided by the cellular network is given by $\bar{R}_c=R_{c}BK_c/K$. Suppose the total bandwidth is $W$, and we \emph{orthogonally} allocate the bandwidth $W$ to the cellular network, denoted by $\alpha W$, and the D2D network, denoted by $(1-\alpha)W$. Then the rate $\alpha \bar{R}_c+(1-\alpha)\bar{R}_k$ is achievable for user $k$ via time-sharing between the two rates provided by cellular and D2D, respectively. By doing so, each user can independently choose either D2D or cellular access at each time, and most of the users whose rates in the D2D network are very close to zero can still be served in the cellular network, so as to alleviate the user fairness.

While the idea of time sharing is very simple and easy to operate, the achievable rate can further be significantly improved by jointly designing user scheduling between D2D and cellular, in particular when we take into account the pathloss and channel fading of the cellular transmission.

\section{Conclusion}

In this paper, we introduced a novel but quite simple binary scheduling scheme based on SINR-threshold scheduling. The scheme was based on treating interference as noise at each receiver and SINR-threshold scheduling with binary power control. In particular, the SINR at each receiver is predicted assuming \emph{all} the other user-pairs are active. This SINR prediction is then compared to an optimal preassigned threshold in order to determine whether or not to activate the individual link. Subsequently, the receiver feeds back this one-bit decision information to its corresponding transmitter, so that the transmitter is either activated with full power or silent for data transmission. As we showed by both analysis and system level evaluations, when compared to the prior schemes such as ITLinQ \cite{navid}, Fair ITLinQ \cite{navid}, FlashLinQ \cite{flash} and the SNR-based method \cite{G}, our proposed scheme outperforms them in terms of sum rate. In addition, the computational complexity of the proposed scheme is lower than ITLinQ and FlashLinQ and on par with the SNR-based method.

\emph{Acknowledgment:} This work is in part supported by NSF grant NETS-1419632 and ONR award N000141612189.


%
%


\bibliographystyle{IEEEtran}
{\footnotesize
\bibliography{globcom_v2_conf}}
\end{document}